\def\rect#1#2{{\vcenter{\vbox{\hrule height.3pt
	    \hbox{\vrule width.3pt height#2truecm \kern#1truecm
	    \vrule width.3pt}
	    \hrule height.3pt}}}}
\def\square{\rect{0.15}{0.15}}
\documentstyle[12pt]{article}
\begin{document}
\thispagestyle{empty}
\begin{center}
\LARGE
Renormalization Group results for lattice surface models.
 ~\\
 ~\\
 ~\\
 ~\\
 ~\\
\normalsize
Emilio N. M. Cirillo, Giuseppe Gonnella\\
\vskip 0.5cm
{\it Dipartimento di Fisica dell'Universit\`{a} di Bari} and\\
{\it Istituto Nazionale di Fisica Nucleare, Sezione di Bari\\
via Amendola 173, I-70126 Bari, Italy}
 ~\\
 ~\\
\normalsize
 ~\\
 ~\\
\end{center}
\begin{abstract}
We study the phase diagram of statistical systems of closed and open
interfaces built on a cubic lattice. Interacting closed interfaces
can be written as Ising models, while open surfaces as $Z(2)$ gauge systems.
When the open surfaces reduce to closed interfaces with  few defects,
also the gauge model can be written as an Ising spin model.
We apply the lower bound renormalization group (LBRG) transformation introduced
by
Kadanoff (Phys. Rev. Lett. \underline {34}, 1005 (1975))
to study the Ising models describing closed and open surfaces
with few defects. In particular, we have studied the Ising-like
transition of self-avoiding surfaces between the random-isotropic phase
and the phase with broken global
symmetry at varying values of the mean curvature.
Our results
are compared with previous numerical work.
The limits of the LBRG transformation in describing regions
of the phase diagram  where not ferromagnetic ground-states are relevant are
also discussed.
\end{abstract}
 ~
transitions);
  11.15.Ha (Lattice gauge theories);
\newpage

\addtolength{\baselineskip}{\baselineskip}

{\bf \S 1. Introduction.}
\par
In this paper we will apply a renormalization group transformation
 to study the phase
diagram of interface models built on a cubic lattice.
Fluid interfaces in $3D$ statistical systems are
subject of much current research \cite{PWN}.
They provide useful descriptions of experimental systems such as mixtures of
oil, water and surfactant, or aqueous solutions of surfactant \cite{GS}.
In ternary mixtures the surfactant forms monolayered
interfaces between oil
and water; in aqueous solutions bilayered membranes  are typical constituents
of biological cells.
The properties of these systems at low surfactant concentrations
are relevant for both practical and theoretical reasons.
For example, in ternary mixtures, a middle phase \cite{deGT}
coexisting with oil-rich and water-rich phases is considered
very appealing for  applications \cite{CML}, due to the very
low surface tension values between the coexisting phases.
{}From a theoretical point of view, dilute interfaces
can be seen as experimental realizations of random surface models where
self-avoidness is the only relevant interaction \cite{P}.
\par
The typical lack of topological constraints on the physical configurations
suggests the use of lattice models to describe ensembles of fluid
surfaces.
First consider the case of closed interfaces without defects such as holes
or seams. Closed interfaces can be described in Ising models
as the boundaries separating
domains of opposite spins, which, in the identification with ternary mixtures,
can represent oil and water. The interfaces
 are built on the dual lattice,
and, for a given spin configuration $\{\sigma_i\}$, have a total area
${\cal S}=\sum_{<ij>}(1-\sigma_i \sigma_j)/2$, where the sum is over all
nearest-neighbour pairs in the original lattice.
Other surface energies can be considered by introducing
further spin interactions. Surfaces where curvature and intersections
\cite{I} are  also weighted can be represented by
a generalization of the Ising model defined  by the
hamiltonian [7--9]
\begin{equation}
 H = {\cal J}_{1} \sum _{<ij>} \sigma_i \sigma_j
+ {\cal J}_{2} \sum _{<<ij>>} \sigma_i \sigma_j
+ {\cal J}_{4} \sum _{^i_l\square^j_k} \sigma_i \sigma_j \sigma_k \sigma_l
\label{eq:1}
\end{equation}

\noindent
where the three sums are respectively over nearest, next-to-the-nearest
neighbours and plaquettes of a cubic  lattice.
Here and in the following of the paper,
 our definitions of hamiltonian will always  include tha factor $-\beta$.
The parameters ${\cal J}_{1}, {\cal J}_{2},{\cal J}_{4}$
can be expressed in terms
of the surface parameters
$\beta_S, \beta_C, \beta_L$ \cite{ccgm} representing respectively
the energy cost for an elementary area (one plaquette
on the dual lattice), for two plaquettes at right angle, and for
four plaquettes with a common bond \cite{II}. A positive $\beta_C$ favours flat
configurations; it corresponds to a mean curvature energy, which has
been proved to be an
useful phenomenological parameter for describing fluid interfaces \cite{CHL}.
The term proportional to $\beta_L$ can mimic the self-avoidness
interaction in the limit $\beta_L\rightarrow\infty$, when surfaces
touching each other along some contour are forbidden. The relations
between spin and surface parameters are:
\begin{equation}
{\cal J}_1={\beta_S+\beta_L\over 2}+\beta_C,\;\;
{\cal J}_2=-{\beta_L\over 8}-{\beta_C\over 4},\;\;
{\cal J}_4=-{\beta_L\over 8}+{\beta_C\over 4}.
\label{eq:2}
\end{equation}

\noindent
The phase diagram of the model (\ref{eq:1}) has been studied by mean-field and
numerical simulations in \cite{ka,ccgm}. It exhibits many
properties relevant for real systems, as also discussed in \cite{cgm}.

If one wishes to consider the effects of defects in fluid interfaces,
ensembles of open surfaces have to be introduced. A simple lattice realization
of open surfaces \cite{HL} is given by the self-dual $Z(2)$ gauge model
\cite{W}.
Here 2-values variables $\{U_{ij}\}$ are defined on the bonds of a cubic
lattice.
One says that the plaquette dual to the bond $<ij>$ is occupied by some surface
if $U_{ij}=-1$; it is not occupied if $U_{ij}=1$. Therefore a given
$\{U_{ij}\}$ configuration corresponds on the dual lattice to a
surface configuration with area
${\cal S}=\sum_{<ij>}(1-U_{ij})/2$. A bond on the dual lattice can be said
to belong to some defect if an odd number of the dual plaquettes sharing that
bond is occupied by some surface. Defects defined in this way
can be counted by
considering the product of $U_{ij}$ over the bonds of each plaquette in the
original lattice. It is easy to recognize that the total length of defects
will correspond to the quantity ${\cal D}=
\sum_{^i_l\square^j_k}(1-U_{ij}U_{jk}U_{kl}U_{li})/2$ \cite{HL}.
Therefore the self-dual $Z(2)$ gauge model with hamiltonian given by

\begin{equation}
H = \beta_M\sum_{<ij>}U_{ij}+\beta_G
\sum_{^i_l\square^j_k}U_{ij}U_{jk}U_{kl}U_{li}
\label{eq:3}
\end{equation}

\noindent
describes open surfaces where area and defects are both weighted.
Self-duality here \cite{W} means that the model is symmetric with respect to
the transformations
\begin{equation}
\beta_M \rightarrow \tilde\beta_G= -{1\over 2} \ln \tanh \beta_M ,
\hskip 0.8cm
\beta_G\rightarrow \tilde\beta_M= -{1\over 2} \ln \tanh \beta_G.
\label{eq:4}
\end{equation}

\noindent
In the parametrization (\ref{eq:3}) a large value of $\beta_M$ favours
configurations with small area, while a large value of $\beta_G$ inhibits
defects.
\par
The phase diagram of the self-dual $Z(2)$ gauge model
has been first analyzed in \cite{W,FS}; it has been studied
by Monte Carlo simulations in \cite{JJJ}. At small values of
$\beta_M$ -  and analogously,  by self-duality, at large values of $\beta_G$
- it can be shown \cite{FS,GH} that the model (\ref{eq:3}) can be expanded as
an Ising spin model with an increasing number of
interactions. For example, at the second order of the expansion at small
$\beta_M$ the model (\ref{eq:3}) can be written as
\begin{equation}
H = {\cal J}_{1}\sum_{<ij>}\sigma_i \sigma_j
+   {\cal J}_{2}\sum_{<<ij>>}\sigma_i \sigma_j
 +  {\cal J}_{4}\sum_{^i_l\square^j_k}\sigma_i\sigma_j\sigma_k\sigma_l
  + {\cal J}_{6}\sum_{\rm cor}\sigma_i\sigma_j\sigma_k\sigma_l
\label{eq:41}
\end{equation}
where the interactions are between nearest neighbours, next-to-the-nearest
neighbours, the 4 spins of a plaquette and the 4 spins of a corner
(see Table 1). The coupling
constants ${\cal J}_{1},{\cal J}_{2},{\cal J}_{4}$ and
${\cal J}_{6}$ can be expressed in terms of the constants $\beta_G$ and
$\beta_M$ as follows
\begin{equation}
\begin{array}{ll}
	      &      \\
{\cal J}_{1} &={\tilde\beta}-4(\tanh\beta_M)^6
                [3\cosh 2{\tilde\beta}\sinh 2{\tilde\beta}
                +(\cosh 2{\tilde\beta})^2\sinh 2{\tilde\beta}]\\
              &      \\
{\cal J}_{2}&= 2(\tanh\beta_M)^6
                [(\sinh 2{\tilde\beta})^2
                +\cosh 2{\tilde\beta}(\sinh 2{\tilde\beta})^2]\\
              &      \\
{\cal J}_{4}&= 2(\tanh\beta_M)^6(\sinh 2{\tilde\beta})^2\\
              &      \\
{\cal J}_{6}&= -{1\over 2}(\tanh\beta_M)^6(\sinh 2{\tilde\beta})^3\\
	      &      \\
\end{array}
\label{eq:42}
\end{equation}
where
\begin{equation}
{\tilde \beta}=-{1\over 2}\ln\tanh\left[\beta_G +(\tanh\beta_M)^4\right].
\label{eq:43}
\end{equation}
The spin representation has the advantage that it can be more easily
studied  \cite{ZD}.
\par
In this paper we will apply the so called lower-bound renormalization
group (LBRG) transformation first proposed by Kadanoff \cite{K}
to study the phase
diagram of the models (\ref{eq:1}) and (\ref{eq:3}) as given in the
approximation (\ref{eq:41}).
The LBRG transformation can be conveniently applied to cases where
all the interaction is in an elementary cell of the lattice, as it is
in the models (\ref{eq:1}) and (\ref{eq:41}). The convenience is appreciable
especially in $D=3$
where other RG transformations would be much more dispendious from a
computational point of view.
\par
The LBRG transformation produces a lower bound to the free-energy
which can be maximized by conveniently
fixing a variational parameter. Its  application
to various models generally gives very accurate estimates  of critical
exponents \cite{BH}.
For example, in the $2D$ Ising model it predicts
the inverse critical temperature
$\beta_{crit}=0.458$ ($\beta_{crit}^{Onsager}=0.4407$), and the exponent
$\nu=0.999$ ($\nu^{Onsager}=1$) \cite{B}.
The drawback of the LBRG transformation is that it preserves the nature
of the ground states only in the ferromagnetic region, so that it can be
reliably applied in a limited region of the phase diagram.
\par
We will describe the LBRG transformation in $\S$2. In $\S$3 we will show
the results obtained by applying the LBRG transformation to the models
(\ref{eq:1}) and (\ref{eq:3},\ref{eq:41}).
In particular, in model (\ref{eq:1}), the self-avoidness limit is examined
for different values of the curvature $\beta_C$. A discussion of our results
will follow in $\S$4.

\vskip 3cm
{\bf \S 2. The LBRG Transformation.}
\par
In this section we will briefly describe the LBRG transformation.
Further details can be found in \cite{K,BH}. Real
space RG  transformations can be generally written as
\begin{equation}
e^{H'(\sigma ',{\cal J}')}=\sum_{\sigma} {\cal P}(\sigma ',\sigma)
e^{H(\sigma ,{\cal J})}.
\label{eq:5}
\end{equation}
Here ${\cal J}$ denotes a set of coupling constants, $\sigma=\{\sigma_1 ,...
\sigma_N\}$ a spin
configuration and $H(\sigma,{\cal J})$ is the hamiltonian to be studied;
the weight function ${\cal P}(\sigma';\sigma)$ defines the
renormalized  hamiltonian $H'(\sigma',{\cal J}')$
with new spin variables $\sigma '_1,...,\sigma '_{N'}$ $(N'<N)$ and coupling
constants
${\cal J}'$. The relation $\sum_{\sigma'}{\cal P}(\sigma';\sigma)=1$
ensures that the total free-energy is unchanged. In the LBRG transformation
\cite{K} the spin  $\sigma'_i$ are defined on the cells like those marked by
a cross in Fig.1; ${\cal P}(\sigma ',\sigma)$ is chosen as the
product over the marked cells of the functions
\begin{equation}
{\hat {\cal P}}(\sigma '_i; \sigma_{i,1} ,...,\sigma_{i,8})=
{\exp [p\sigma '_i(\sigma_{i,1} + ... +\sigma_{i,8})]\over
 2\cosh [p(\sigma_{i,1} + ... +\sigma_{i,8})]}\;\; \forall i=1...N'
\;\; ,
\label{eq:6}
\end{equation}
with $p$ a real parameter and $\sigma_{i,1},...\sigma_{i,8}$
the original spins at the
vertices of the cube $i$.
\par
If the original hamiltonian can be written as $H(\sigma,{\cal J})=
\sum {\hat H}(\sigma,{\cal K})$, where the sum
is over the elementary cubes of the lattice and ${\cal K}$ is the set of
couplings normalized to a single cell \cite{III}, a convenient {\it moving} of
interactions and factors of (\ref{eq:6}) will give a new hamiltonian
with all the interaction still in a single cell.
The renormalized cell hamiltonian ${\hat H}'(\sigma',{\cal K}')$ is given by
\begin{equation}
\exp [{\hat H}'(\sigma '_1,...,\sigma '_8;{\cal K}')]=
\sum_{\sigma_1,...,\sigma_8}
{\exp [p(\sigma '_1\sigma_1+...+\sigma '_8\sigma_8)+8 {\hat H}(\sigma_1,...,
\sigma_8;{\cal K})]\over
 2\cosh [p(\sigma_1+...+\sigma_8)]}\;\; .
\label{eq:7}
\end{equation}
\par
Since we are interested in studying the phase diagrams of
the Ising models (\ref{eq:1}) and
(\ref{eq:41}) where only even interactions appear, it will be  sufficient to
consider the transformation of the 14 even couplings (see Table I) which can
be defined on a cell of a $3D$ cubic lattice.
One of these couplings is a pure constant, we denote it by ${\cal K}_0$; the
others are denoted by ${\cal K}_i$ $i=1,...,13$.
\par
After some algebra one gets from (\ref{eq:7}) the recursion laws
\begin{equation}
\left\{
\begin{array}{ll}
{\cal K}'_0=& 8{\cal K}_0+\Phi_0 (p;{\cal K}_1,...,{\cal K}_{13})\\
{\cal K}'_i=& \Phi_i (p;{\cal K}_1,...{\cal K}_{13})\;\; \forall i=1,...,13\\
\end{array}
\right.
\label{eq:8}
\end{equation}
where $\Phi_0,...,\Phi_{13}$ are analytic functions. The critical properties of
the system can be then related to the behaviour of the recursion laws close to
their fixed points.
\par
The variational nature of the interaction-moving operation
was first observed by Kadanoff \cite{K1}. A
lower bound $f^*(p)$ to the free-energy per site
can be calculated by
\begin{equation}
f^*(p)=-\lim_{n\to\infty} {{\cal K}_0^{(n)}\over 8^n}\;\; ;
\label{eq:9}
\end{equation}
where ${\cal K}_0^{(n)}$ is the value of ${\cal K}_0$ after $n$ applications
of the LBRG transformation \cite{K}. Following the prescription of \cite{K},
the parameter $p$ will be fixed by maximizing the function $f^*(p)$ starting
the iterations from the fixed point hamiltonian with ${\cal K}_0=0$.

\vskip 3cm
{\bf \S 3.  Results.}
\par
{\it Closed interfaces - model (\ref{eq:1})}.
The LBRG transformation is here applied
to calculate the ferromagnetic-paramagnetic (F-P) transition surface in the
space ${\cal J}_1, {\cal J}_2, {\cal J}_4$. For completeness, results
concerning
other transitions, related to not ferromagnetic ordering,
will be
also given. These results have to be considered with cautions since
the LBRG transformation, as defined in \S2,
does not take correctly in account the structure of not
ferromagnetic ground states.
\par
The  value of $p$ maximizing the critical fixed point
free-energy  on the F-P surface is $p^*_c=0.40354$.
In Table 1
fixed points related to the F,P and AF (antiferromagnetic)
phases are reported
for the value $p=p^*_c$.
\par
The fixed points $\scriptstyle (F)$, $\scriptstyle (P)$ and $\scriptstyle (C)$
are respectively
the low-temperature ferromagnetic, the high
temperature and the F-P critical fixed points. The LBRG transformation has
been already
applied
for calculating the exponents of the $3D$ Ising model
in \cite{K},
where the optimal value found for
$p^*$ is  $p=0.40343$. We do not understand the
reasons of the discrepancy with our result.
The critical fixed point at $p=0.40343$
is  reported in the caption of Table 1.
At $p=p^*_c$ the values of the inverse Ising critical temperature and of
the exponent $\nu$ are respectively
$\beta_{crit}^I=0.23925$   and $\nu=0.6288$.
The corresponding values at $p=0.40343$ are
$\beta_{crit}^I=0.23923$  and $\nu=0.6290$;
the  best estimates \cite{FL}
are $\beta_{crit}^I=0.22165$ and $\nu=0.6289\pm 0.0008$. The fixed
point $\scriptstyle (C)$ is symmetric in the sense that all  the
2-spin, the 4-spin,
etc. interactions  are equal. This symmetry was assumed in \cite{K}, while
here we consider recursions in the whole space of couplings.
This situation can be compared
with the results obtained by applying  the LBRG transformation
to the $2D$ Ising model \cite{B}.
In $D=2$ the symmetric  fixed point has two relevant
eigenvalues  with an
 eigenvector pointing
outside the symmetric subspace on the critical surface. Therefore in $D=2$,
differently from the $3D$ case,
the symmetric critical fixed point, which is found to
maximize the free-energy,  cannot be reached starting from  non symmetric
interactions \cite{B}.
\par
The fixed points $\scriptstyle (AF)$ and $\scriptstyle (AC)$ are the
antiferromagnetic counterparts of the fixed points $\scriptstyle (F)$ and
$\scriptstyle (C)$.
The fixed point $\scriptstyle (AC)$ is on the transition surface between
the AF and the P phases; its exponent is $\nu=0.6349$. This surface
intersects the surface ${\cal F}$
limiting the F phase  at positive ${\cal J}_1$ as shown in Fig. 2.
The model (\ref{eq:1}) exhibits the exact symmetry
${\cal J}_{1}
\rightarrow - {\cal J}_{1}$ \cite{ccgm}.  This symmetry
is not respected in Fig. 2.
However, we observe
that  simple block transformations would completely miss the F-AF transition
in the $2D$ version of the model (\ref{eq:1}) \cite{NN}.
On the surface between the F and the AF phases we find the
fixed point
$\scriptstyle (D)$; it has one
relevant eigenvalue given at $p=p^*_c$ by
$\lambda=2^{y_D}$ with $y_D=2.72454$. We interpret the point $\scriptstyle (D)$
as a discontinuity fixed point related to the F-AF first-order transition,
which should be characterized by the value $y_D=D=3$ \cite{NN1}.
If we maximize the
free energy with respect to the discontinuity fixed point, we get $y_D=1.78$
at
$p=0.31$, which is the lowest value for which the discontinuity fixed point
exists. The fact that this
 result is worse than the one obtained at $p=p^*_c$ can be explained by saying
that the LBRG transformation does not give good results when not
ferromagnetic ground-states are involved.

Numerical simulations of \cite{ka,ccgm} show the existence of a
line of tricritical points on the F-P transition surface
 close to the ${\cal J}_1=0$
 plane; this line, at decreasing values of ${\cal J}_4$, ends in a Baxter
point.
Due to the limits
of applicability
of the LBRG transformation  at small ${\cal J}_1$,
we cannot give reliable predictions on the structure of the phase diagram
in the region
where the F-P and the AF-P surfaces meet.
However, we have also studied the RG recursions on
the line ${\cal L}$ separating the domains
of attraction of the fixed points
$\scriptstyle (C)$ and $\scriptstyle (D)$ on the surface ${\cal F}$.
On the line ${\cal L}$, which is very close to the intersection
of the AF-P with the
${\cal F}$ surface,
we find a fixed point $\scriptstyle (L)$ with two relevant eigenvalues,
which annihilates with the discontinuity fixed point for
$p<0.31$.
The free-energy  of this fixed point is maximum at $p$ very close  to
$p^*_c$, where the exponents of the two relevant eigenvalues are
$y_1=1.59347$ and $y_2=0.09564$ \cite{CC}.
The largest not relevant eigenvalue is $\lambda=0.90395$.
The fixed point $\scriptstyle (L)$ is also reported in Table 1; it can be seen
that it is very close to the fixed point $\scriptstyle (C)$. A realistic
discussion of the phase diagram in the plane ${\cal J }_1=0$, where the F-P and
the AF-P surfaces should meet, is given in \cite{jmj}.
\par
In Fig.3 the phase diagram is shown in the particular case
${\cal J}_4={\cal J}_2$, which means $\beta_C=0$ in the surface representation.
The paramagnetic phase, in accord with Monte Carlo results and differently from
what comes out from mean field approximation \cite{ccgm},
extends  at positive
${\cal J}_1$ towards zero temperature.
This is related to the high degeneracy of the ground states in this region
\cite{ccgm}.
\par
A different representation of the phase diagram can be given in terms of the
surface parameters $\beta_L, \beta_S$ and $\beta_C$  (see (\ref{eq:2})).
In Fig.4 the F-P-AF transitions are shown in the plane $\beta_L$,$\beta_S$ for
different values of the curvature $\beta_C$. The F phase, at large values of
$\beta_S$, describes configurations
with  diluted small surfaces. By decreasing the value of $\beta_S$,
area is favoured to increase and, at the percolation threshold,
 interfaces
invade the system. However, it is still possible to distinguish between an
{\it inside} volume wrapped up in interfaces and a different
 {\it outside} volume. By decreasing furtherly the value of $\beta_S$,
if $\beta_L$ is sufficiently large,
at the Ising-like F-P transition, a {\it random isotropic} \cite{HL1}
phase is stable and the symmetry of the hamiltonian
between inside and outside is restored.
The AF phase can be intended
as a droplet crystal.
The limit $\beta_L \rightarrow \infty$
describes  a gas of self-avoiding surfaces and is particularly relevant
for physics.
In Table 2 the critical values of  $\beta_S$
for self-avoiding surfaces are reported at different values of the
curvature and compared with results from simulations.

\vskip 0.7cm
{\it Open interfaces - model (\ref{eq:3},\ref{eq:41})}.
The gauge model (\ref{eq:3}) at $\beta_M=0$ is dual to the $3D$ Ising model
\cite{W}. At
small $\beta_M$ it can be expanded on the dual lattice as an Ising
model with many interactions. At the
second order of this expansion the
gauge model is mapped onto the model (\ref{eq:41}).
The LBRG transformation has been
applied to study the F-P transition in the
model (\ref{eq:41}). Then the results have
been reported by formulas (\ref{eq:42},\ref{eq:43}) in the
plane $\beta_M\beta_G$, as shown in Fig.5.
At small $\beta_M$ the critical line starts from the $\beta_G$ axis at
$\beta_G=-{1\over 2}\ln\tanh\beta_{crit}^I$, with $\beta_{crit}^I=0.23926$.
The continuous line in Fig.5 is the self-dual line, which is the line mapped
onto
itself by transformations (\ref{eq:4}), with
respect to which the phase diagram has to
be symmetric. Then  the
critical line at small $\beta_M$ is mapped by eqs. (\ref{eq:4})
in the region at large $\beta_G$. The two lines meet at the point
$\beta_M=0.241,
\beta_G=0.719$ on the self-dual line.
In the phase diagram found by numerical simulations \cite{JJJ}, the two
lines starting at $\beta_M=0$ and at $\beta_G=\infty$ become first-order
at tricritical points before meeting on the self-dual line.
There, at a triple point, an other first-order
line comes out towards greater values of $\beta_M$ on the self-dual line.
This first-order line ends with  a critical point at
finite and positive values of $\beta_G$ and $\beta_M$. As discussed in
\cite{GH}, an interesting aspect of the expansion (\ref{eq:41}) is that the
4-spin interaction
terms are expected to give tricritical points.
However,  a mean-field
approximation  of the model (\ref{eq:41}) \cite{GH}
gives tricritical points quite far beyond the triple point.
Also our calculations suggest that the transition lines are continuous
on the parts drawn in Fig.5,
which correspond to the critical F-P
transition in the Ising representation.
Therefore, the relevance of the 4-spin interaction in (\ref{eq:41})
is probably not sufficient to explain alone the existence of the
tricritical points found in simulations.
These results will be further commented in the  next section.

\vskip 3cm
{\bf \S 4. Discussions and conclusions.}
\par
We have applied the LBRG transformation to study the phase
diagram of Ising models describing closed and open interfaces.
Interacting closed interfaces
can be naturally expressed as an Ising model,
while open surfaces, originally written as a gauge model with statistical
variables on the bonds, can be mapped on  Ising models only in extreme
regions of the phase diagram.
At large $\beta_G$, the gauge model
describes the interesting physical situation of {\it almost}-closed
surfaces with few defects.
\par
First consider the model (\ref{eq:1}) of closed interfaces.
Results concerning
the transition on the nearest-neighbour axis are in good agreement with
previous known
results. Also the value of the Ising exponent $\nu=0.6289$
is in excellent agreement with other numerical work.
We expect that the critical surface has been found with a good approximation
in the region close to the nearest-neighbour axis.
Results regarding the interesting case of self-avoiding surfaces have
been reported in Table 2.
\par
Problems arise when the LBRG transformation is applied to study regions
of the phase diagram where
ordered not ferromagnetic configurations are relevant.
In particular, the LBRG transformation does not take
into account the ${\cal J}_1\rightarrow - {\cal J}_1$ symmetry
of the model (\ref{eq:1})
which should give at low temperatures a first-order F-AF transition at ${\cal
J}_1=0$. We find this first-order transition, but not at ${\cal J}_1=0$ (see
Figs.2,3).
Moreover,  our results cannot reliably describe the region close
to the line where the F-P and the AF-P surfaces meet, which should be on the
plane ${\cal J}_1=0$.
However, for completeness, we have also given results concerning this region.
\par
The model (\ref{eq:1}) has been largely studied in
$D=2$ \cite{BA}, where
RG transformations taking correctly into account the ground-state structure
have been considered giving the expected topology of the phase
diagram \cite{jmj,H}.
We have tried to generalize the LBRG transformation in
order to
take into account the existence of antiferromagnetic ground states. Then we
have considered a weight function ${\cal P}(\sigma',\sigma)$
distinguishing between
spins of different sublattices.
For each cell marked by a cross in Fig.1
the spin $\sigma'$ is coupled only to the four spins of one original
sublattice (see eq. (\ref{eq:6}) and Fig.1), in such a way that two
nearest neighbouring spins $\sigma '$ are coupled to the spins $\sigma$
of different sublattices. Then
all the interaction is {\it moved} into the dark grey cells of Fig.1 and a RG
transformation analogous
to eq.(\ref{eq:7}) can be written in such a way to get a homogeneous
hamiltonian with the same expression for any elementary cell.
By this procedure we have obtained phase
diagrams which exhibit the symmetry ${\cal J}_1 \rightarrow - {\cal J}_1$,
but with a rather poor precision for the critical temperature on the
nearest-neigbour axis and for the exponent $\nu$. Moreover the
tricritical points numerically found \cite{ccgm}
close to the plane ${\cal J}_1=0$ are not
obtained by this transformation. Therefore a complete RG study of the phase
diagram of the model (\ref{eq:1}) in $D=3$ is still an open question.
\par
In Fig.5 we have presented the phase diagram of the self-dual $Z(2)$ gauge
model
found by applying the LBRG
transformation to the model (\ref{eq:41}).
Our estimation of the critical lines is reliable especially
in the region of validity of the expansion (\ref{eq:41}), that is at
small $\beta_M$ and, by duality,  at large $\beta_G$, close to the points
where the model can be written as an Ising model with only nearest-neighbour
interaction.
Numerical simulations \cite{JJJ} predict that these lines become first-order
before meeting on the self-dual line. By our methods, we cannot predict
such a behaviour.  Indeed, our results suggest that the transition line
remain continuous for a long part beyond the self-dual line. Therefore, even
if tricritical points could arise in model (\ref{eq:41}), that expansion
is probably not useful to discuss the phase diagram of the model
(\ref{eq:3}) close to the self-dual line, for which other methods are needed.
In conclusions, provided all the
discussed limitations, we can say that the application of the LBRG
transformation to spin models describing lattice interfaces gives, in a
relatively simple way, phase diagrams in many parameter spaces
which are quite accurate especially
in  the region close to the nearest-neighbour axis.
\vskip 2cm
{\bf Acknowledgements}

One of the authors (G. G.) thanks Prof. Attilio Stella for an useful discussion
about the subject of this work. We thank Mr. Alexis Pompili for having drawn
Fig.1.

\newpage
\Large
\begin{center}
{\bf Table Captions}
\end{center}
\normalsize
\addtolength{\baselineskip}{\baselineskip}
\par\noindent
Table 1.
\par\noindent
The coordinates of the fixed point related to the F-AF-P transitions
are reported for the value $p=p^*_c$ which maximizes the free energy
of the critical F-P fixed point $\scriptstyle (C)$. The other symbols
$\scriptstyle (F),\; (AF),\; (P),\; (AC),\; (D)$ and $\scriptstyle (L)$ denote
respectively the low-temperature ferromagnetic and antiferromagnetic fixed
points, the high-temperature, the critical AF-P, the discontinuity fixed point
between the AF and F phases, and the fixed point on the manifold separating the
domains of attraction of the fixed points $\scriptstyle (C)$ and
$\scriptstyle (D)$ (on the hypersurface limiting the F phase).
The two squares on the
left represent two parallel faces of an elementary cube of the lattice. The
dots represent the spins taking part in a given interaction. In \cite{K} the
value of $p$ maximizing the critical F-P fixed point
free energy has been found to be
$p=0.40343$. This fixed point is symmetric (in the sense explained in the main
text) and the 2--spins, 4--spins, 6--spins and 8--spins coordinates are
respectively $0.02097$, $1.96\times 10^{-4}$, $-7.69\times 10^{-5}$ and
$2.15\times 10^{-5}$.
\par\noindent
Table 2.
\par\noindent
The critical values of $\beta_S$ in the self-avoidness limit for different
values of $\beta_C$. The Monte Carlo results are taken from Refs
\cite{ka} and \cite{ccgm}.

\newpage
\Large
\begin{center}
{\bf Figure Captions}
\end{center}
\normalsize
\addtolength{\baselineskip}{\baselineskip}
\par\noindent
Fig.1: A $2D$ representation of the LBRG transformation. The crosses indicate
the spins $\sigma '$; the squares are the original spins $\sigma$.
The $\sigma$--dependent terms in the hamiltonian and in the weight function are
moved into the grey squares. Full and empty squares
represent spins of the two original sublattices. In the variant of the LBRG
transformation described in \S 4 the $\sigma$--dependent terms are moved into
the dark grey cells of the lattice.
\par\noindent
Fig.2: The phase diagram obtained by applying the LBRG transformation to the
model (\ref{eq:1}). The horizontal surface at positive ${\cal J}_2$ separates
the F phase at positive ${\cal J}_1$ from the AF phase. At lower values of
${\cal J}_2$ the F and the AF phases are separated by the P phase.
\par\noindent
Fig.3: The phase diagram of the model (\ref{eq:1}) with ${\cal J}_2={\cal
J}_4$.
Fixed points of Table 1 are also reported.
\par\noindent
Fig.4: The phase diagram of the model (\ref{eq:1}) in terms of the surface
parameters $\beta_S$ and $\beta_L$. The different curves refer from the right
to the left respectively to the values of $\beta_C=-0.2,-0.1,-0.04,0$. The
symbols F, AF and RI denote respectively the ferromagnetic, the
antiferromagnetic and the random isotropic or paramagnetic phase.
\par\noindent
Fig.5: The phase diagram of the self-dual $Z(2)$ gauge model. The continuous
line is the self-dual line. The dashed lines are critical lines found by
applying the LBRG transformation to the model (\ref{eq:41}).

\newpage
\thispagestyle{empty}
\begin{center}
{\bf Table 1}
\end{center}

\begin{center}
\begin{tabular}{llccccccc} \hline\hline
       &
       &$\scriptstyle {(F)}$
       &$\scriptstyle {(AF)}$
       &$\scriptstyle {(P)}$
       &$\scriptstyle {(C)}$
       &$\scriptstyle {(D)}$
       &$\scriptstyle {(AC)}$
       &$\scriptstyle {(L)}$\\ \hline
$^{\bullet}\hskip -0.12truecm\square2^{\hskip -0.1truecm\bullet}$
\hskip 0.12 truecm
$^{\phantom\bullet}\hskip -0.12truecm\square2^{\hskip
-0.1truecm{\phantom\bullet}}$
&${\cal K}_1^*$& ${\phantom {-}}0.09371$&$-0.09447$& ${\phantom {-}}0.04205$
& ${\phantom {-}}0.02096$& ${\phantom {-}}0.01720$
&$-1.19\times 10^{-3}$&${\phantom {-}}0.02010$\\
$^{\bullet}\hskip -0.12truecm\square2_{\hskip -0.1truecm\bullet}$
\hskip 0.12 truecm
$^{\phantom\bullet}\hskip -0.12truecm\square2^{\hskip
-0.1truecm{\phantom\bullet}}$
&${\cal K}_2^*$& ${\phantom {-}}0.09371$& ${\phantom {-}}0.09447$&$-0.04231$&
${\phantom {-}}0.02096$& ${\phantom {-}}0.11550$
&$-1.19\times 10^{-3}$&${\phantom {-}}0.02293$\\
$^{\bullet}\hskip -0.12truecm\square2^{\hskip -0.1truecm{\phantom\bullet}}$
\hskip 0.12 truecm
$^{\phantom\bullet}\hskip -0.12truecm\square2_{\hskip -0.1truecm\bullet}$
&${\cal K}_3^*$& ${\phantom {-}}0.09371$&$-0.09447$&$-0.12974$&
${\phantom {-}}0.02096$& ${\phantom {-}}0.01720$
&$-0.11342$&${\phantom {-}}0.01791$\\ \hline
$^{\bullet}_{\bullet}\hskip -0.12truecm\square2^{\hskip
-0.1truecm\bullet}_{\hskip -0.1truecm\bullet}$
\hskip 0.12 truecm
$^{\phantom\bullet}\hskip -0.12truecm\square2^{\hskip
-0.1truecm{\phantom\bullet}}$
&${\cal K}_4^*$&$-9.99\times 10^{-3}$&$-0.01018$&
${\phantom {-}}6.54\times 10^{-3}$
& ${\phantom {-}}1.97\times 10^{-4}$&$-2.65\times 10^{-4}$&$-6.94\times
10^{-7}$
& ${\phantom {-}}1.96\times 10^{-4}$\\
$^{\bullet}_{\bullet}\hskip -0.12truecm\square2^{\hskip -0.1truecm\bullet}$
\hskip 0.15 truecm
$_{\bullet}\hskip -0.12truecm\square2$
&${\cal K}_5^*$&$-9.99\times 10^{-3}$&$-0.01018$&$-2.73\times 10^{-3}$
& ${\phantom {-}}1.97\times 10^{-4}$&$-2.65\times 10^{-4}$&
${\phantom {-}}1.23\times 10^{-5}$
& ${\phantom {-}}2.31\times 10^{-4}$\\
$^{\bullet}_{\bullet}\hskip -0.12truecm\square2^{\hskip -0.1truecm\bullet}$
\hskip 0.15 truecm
$^{\bullet}\hskip -0.12truecm\square2$
&${\cal K}_6^*$&$-9.99\times 10^{-3}$& ${\phantom {-}}0.01018$&
$-6.55\times 10^{-3}$
& ${\phantom {-}}1.97\times 10^{-4}$&$-3.07\times 10^{-3}$&$-6.95\times
10^{-7}$
& ${\phantom {-}}1.48\times 10^{-4}$\\
$^{\bullet}\hskip -0.12truecm\square2_{\hskip -0.1truecm\bullet}$
\hskip 0.15 truecm
$^{\bullet}\hskip -0.12truecm\square2^{\hskip -0.1truecm\bullet}$
&${\cal K}_7^*$&$-9.99\times 10^{-3}$& ${\phantom {-}}0.01018$&
${\phantom {-}}2.72\times 10^{-3}$
& ${\phantom {-}}1.97\times 10^{-4}$&$-3.07\times 10^{-3}$&
${\phantom {-}}1.23\times 10^{-5}$
& ${\phantom {-}}1.82\times 10^{-4}$\\
$^{\bullet}\hskip -0.12truecm\square2^{\hskip -0.1truecm\bullet}$
\hskip 0.15 truecm
$_{\bullet}\hskip -0.12truecm\square2_{\hskip -0.1truecm\bullet}$
&${\cal K}_8^*$&$-9.99\times 10^{-3}$&$-0.01018$&$-4.63\times 10^{-3}$
& ${\phantom {-}}1.97\times 10^{-4}$&$-2.65\times 10^{-4}$&
${\phantom {-}}1.65\times 10^{-4}$
& ${\phantom {-}}2.67\times 10^{-4}$\\
$^{\bullet}\hskip -0.12truecm\square2_{\hskip -0.1truecm\bullet}$
\hskip 0.15 truecm
$_{\bullet}\hskip -0.12truecm\square2^{\hskip -0.1truecm\bullet}$
&${\cal K}_9^*$&$-9.99\times 10^{-3}$&$-0.01018$&
${\phantom {-}}6.59\times 10^{-3}$
& ${\phantom {-}}1.97\times 10^{-4}$&$-2.51\times 10^{-2}$&$-6.94\times
10^{-7}$
& ${\phantom {-}}1.73\times 10^{-5}$\\ \hline
$^{\bullet}\hskip -0.12truecm\square2^{\hskip -0.1truecm\bullet}$
\hskip 0.15 truecm
$^{\bullet}_{\bullet}\hskip -0.12truecm\square2^{\hskip
-0.1truecm\bullet}_{\hskip -0.1truecm\bullet}$
&${\cal K}_{10}^*$& ${\phantom {-}}4.68\times 10^{-3}$&$-4.76\times 10^{-3}$&
$-8.98\times 10^{-4}$&$-7.72\times 10^{-5}$& ${\phantom {-}}5.59\times 10^{-4}$
&$-2.37\times 10^{-7}$&$-7.84\times 10^{-5}$\\
$^{\bullet}\hskip -0.12truecm\square2_{\hskip -0.1truecm\bullet}$
\hskip 0.15 truecm
$^{\bullet}_{\bullet}\hskip -0.12truecm\square2^{\hskip
-0.1truecm\bullet}_{\hskip -0.1truecm\bullet}$
&${\cal K}_{11}^*$& ${\phantom {-}}4.68\times 10^{-3}$&
${\phantom {-}}4.76\times 10^{-3}$&
${\phantom {-}}9.00\times 10^{-4}$&$-7.72\times 10^{-5}$&
${\phantom {-}}2.82\times 10^{-4}$&$-2.37\times 10^{-7}$
&$-7.56\times 10^{-5}$\\
$^{\phantom\bullet}_{\bullet}\hskip -0.12truecm\square2^{\hskip
-0.1truecm\bullet}_{\hskip -0.1truecm\bullet}$
\hskip 0.15 truecm
$^{\bullet}_{\bullet}\hskip -0.12truecm\square2^{\hskip -0.1truecm\bullet}$
&${\cal K}_{12}^*$& ${\phantom {-}}4.68\times 10^{-3}$&$-4.76\times 10^{-3}$&
${\phantom {-}}1.08\times 10^{-3}$&$-7.72\times 10^{-5}$&
${\phantom {-}}5.59\times 10^{-4}$&$-6.47\times 10^{-6}$
&$-8.06\times 10^{-5}$\\ \hline
$^{\bullet}_{\bullet}\hskip -0.12truecm\square2^{\hskip
-0.1truecm\bullet}_{\hskip -0.1truecm\bullet}$
\hskip 0.15 truecm
$^{\bullet}_{\bullet}\hskip -0.12truecm\square2^{\hskip
-0.1truecm\bullet}_{\hskip -0.1truecm\bullet}$
&${\cal K}_{13}^*$&$-7.69\times 10^{-3}$&$-7.88\times 10^{-3}$&
${\phantom {-}}5.96\times 10^{-7}$& ${\phantom {-}}2.15\times 10^{-5}$&
$-3.04\times 10^{-4}$& ${\phantom {-}}3.42\times 10^{-7}$
& ${\phantom {-}}2.15\times 10^{-5}$\\
\hline\hline
\end{tabular}
\vskip 1 truecm
\end{center}

\newpage
\thispagestyle{empty}
\begin{center}
{\bf Table 2}
\end{center}
\vskip 1 truecm
\begin{center}
\begin{tabular}{ccccccc} \hline\hline
 & & & & & & \\
 & &$\beta_c$& & & & \\
 & & & & & & \\
 & &0.04&  0  &-0.04& -0.1& -0.2\\ \hline
 & & & & & & \\
$\beta_s$&LBRG&0.446&0.472&0.504&0.560&0.674\\
 & & & & & & \\
 &Monte Carlo&-&$0.353^{[8]}$&$0.360^{[9]}$&$0.470^{[9]}$&$0.570^{[9]}$\\
 & & & & & & \\ \hline\hline
\end{tabular}
\end{center}


\begin{thebibliography}{99}
\bibitem{II}The ambiguities mentioned in \cite{I} are practically overcome by
saying that there is an energy cost $\beta_L$ for each elementary
intersection whatever is the surface interpretation of the intersection.
\bibitem{III}For example, if ${\cal J}_1$ denotes the original nearest
neighbour coupling, then ${\cal K}_1={\cal J}_1/4$, since each bond
belongs to four different cells of the $3D$ lattice.
\end{thebibliography}
\end{document}